\documentclass[%
 reprint,
 amsmath,amssymb,
 aps,
]{revtex4-1}

\usepackage{graphicx}
\usepackage{dcolumn}
\usepackage{bm}
\usepackage{hyperref}
\usepackage[mathlines]{lineno}
\usepackage{amsmath,bm}
\usepackage{times}
\usepackage{fouriernc}

\newcommand{\bftau}{\mbox{\boldmath$\tau$}}
\newcommand{\bfpsi}{\mbox{\boldmath$\psi$}}

\begin{document}

\preprint{APS/123-QED}



\title{Structural and Magnetic Phase Transitions in CeCu$_{6-x}T_x$ ($T$ = Ag, Pd)}

\author{L. Poudel$^{1,2}$, C. de la Cruz$^{2}$, E. A. Payzant$^{3}$, A. F. May$^{4}$, M. Koehler$^{5}$, V. O. Garlea$^2$, A. E. Taylor$^{2}$, D. S. Parker$^4$, H. B. Cao$^{2}$, M. A. McGuire$^{4}$, W. Tian$^{2}$, M. Matsuda$^{2}$, H. Jeen$^{4,6}$, H. N. Lee$^{4}$, T. Hong$^{2}$, S. Calder$^{2}$, H. D. Zhou$^{1}$, M. D. Lumsden$^{2}$, V. Keppens$^{5}$, D. Mandrus$^{1,4,5}$ and A. D. Christianson$^{1,2}$}

\affiliation{
$^1$Department of Physics \& Astronomy, University of Tennessee, Knoxville, TN-37966, USA\\
$^2$Quantum Condensed Matter Division, Oak Ridge National Laboratory, Oak Ridge, TN-37831, USA\\
$^3$Chemical \& Engineering Materials Division, Oak Ridge National Laboratory, Oak Ridge, TN 37831, USA\\
$^4$Materials Science \& Technology Division, Oak Ridge National Laboratory, Oak Ridge, TN-37831, USA\\
$^5$Department of Material Science \& Engineering, University of Tennessee, Knoxville, TN-37966, USA\\
$^6$Department of Physics, Pusan National University, Busan 609-735, Korea
}%


\date{\today}

\begin{abstract}
The structural and the magnetic properties of CeCu$_{6-x}$Ag$_x$ (0 $\leq$ $x$ $\leq$ 0.85) and CeCu$_{6-x}$Pd$_x$ (0 $\leq$ $x$ $\leq$ 0.4) have been studied using neutron diffraction, resonant ultrasound spectroscopy (RUS), x-ray diffraction measurements and first principles calculations. The structural and magnetic phase diagrams of CeCu$_{6-x}$Ag$_x$ and CeCu$_{6-x}$Pd$_x$ as a function of Ag/Pd composition are reported. The end member, CeCu$_6$, undergoes a structural phase transition from an orthorhombic ($Pnma$) to a monoclinic ($P2_1/c$) phase at 240 K. In CeCu$_{6-x}$Ag$_x$, the structural phase transition temperature (${T_{s}}$) decreases linearly with Ag concentration and extrapolates to zero at $x_{S}$ $\approx$ 0.1. The structural transition in CeCu$_{6-x}$Pd$_x$ remains unperturbed with Pd substitution within the range of our study. The lattice constant $b$ slightly decreases with Ag/Pd doping, whereas, $a$ and $c$ increase with an overall increase in the unit cell volume. Both systems, CeCu$_{6-x}$Ag$_x$  and CeCu$_{6-x}$Pd$_x$, exhibit a magnetic quantum critical point (QCP), at $x$ $\approx$ 0.2 and $x$ $\approx$ 0.05 respectively. Near the QCP, long range antiferromagnetic ordering takes place at an incommensurate wave vector ($\delta_1$ 0 $\delta_2$) where $\delta_1 \sim 0.62$, $\delta_2 \sim 0.25$, $x$ = 0.125 for CeCu$_{6-x}$Pd$_x$  and $\delta_1 \sim 0.64$, $\delta_2 \sim 0.3$, $x$ = 0.3 for CeCu$_{6-x}$Ag$_x$. The magnetic structure consists of an amplitude modulation of the Ce-moments which are aligned along the $c$-axis of the orthorhombic unit cell. 
\end{abstract}

\maketitle

\section{\label{sec:level1}Introduction}
Understanding the nature of a quantum critical point (QCP) remains one of the most topical questions in condensed matter physics. The conventional model of a metallic QCP proposed by Hertz, Millis, and Moriya (HMM) describes the nature of the critical phenomena within the confines of an instability of a spin-density-wave\cite{hertz1976quantum,MillisPRB,moriya1985spin}. Many systems near a QCP are consistent with the description of the HMM model\cite{knafo2009antiferromagnetic,RevModPhys.79.1015,Kadowaki2004}. For example: divergence of the Gr\"uneisen ratio is observed in CeNi$_2$Ge$_2$ with the exponent $x$ = 1\cite{Kuchler2003}; the heat capacity in Ce(Ni$_{1-x}$Pd$_x$)$_2$Ge$_2$ and CeCu$_2$Si$_2$ diverges with the relation $\gamma = \gamma_0 - \alpha T^{1/2}$\cite{Kuchler2003,Gegenwart98,CWwang}; the resistivity at the ferromagnetic QCP of Ni$_x$Pd$_{1-x}$ obeys power law relation $\rho = \rho_0 + a T^{5/3}$\cite{Nicklas99}. However, the HMM model is not sufficient to explain many properties observed in a number of systems near a QCP\cite{Friedemann17082010,PhysRevLett.75.725, coleman2012heavy,si2001locally,gegenwart2008quantum,paschen2004hall}, and a substantial subset of these systems are interpreted as hosting a ``Local QCP"\cite{si2001locally,gegenwart2008quantum,paschen2004hall,Jiao20012015,coleman2001}, where the breakdown of the Kondo-screening leaves the local-moments free to form a magnetic ground state. 

The well-known heavy fermion system CeCu$_{6-x}$Au$_x$ (x $\approx$ 0.1) is often considered as a prototypical example of a local QCP\cite{schroder2000onset,lohneysen2006rare,schroder1998scaling,lohneysen2007magnetic,Stockert1999376}. The nature of the spin fluctuation spectrum of CeCu$_{5.9}$Au$_{0.1}$ is peculiar and is not consistent with the conventional theory that successfully describes several aspects of many Ce-based heavy fermion materials\cite{schroder2000onset,lohneysen2006rare,schroder1998scaling,lohneysen2007magnetic,Stockert1999376}. Inelastic neutron scattering measurements of  CeCu$_{5.9}$Au$_{0.1}$ show that the imaginary part of the dynamic susceptibility at the QCP exhibits an $E/T$ scaling relation $\chi^{\prime \prime}({\bf Q}, E) = T^{-\alpha} f(E/T)$ with an anomalous value of the scaling exponent $\alpha \approx$ 0.75\cite{schroder2000onset,schroder1998scaling,stockert2010quantum,Löhneysen2000480}. The scaling relation as well as the logarithmic divergence of $C/T$ with temperature in CeCu$_{6-x}$Au$_x$ are in accord with the behavior expected for a local QCP.   

While great attention has been given to the evolution of magnetic properties with Au doping into CeCu$_6$, less attention has been paid to the evolution of the structural properties. The CeCu$_{6-x}$Au$_x$ system exhibits a structural phase transition from orthorhombic ($Pnma$) to monoclinic ($P2_1/c$) that can be tuned by pressure or chemical doping\cite{grube1999suppression,robinson2006quantum}. Previous studies have reported different values of the structural transition temperature (${T_{s}}$) for the end member, CeCu$_6$, but are all within the range 168 K - 230 K\cite{grube1999suppression,robinson2006quantum,Vrtis1986489,suzukisoft,Goto1987309}. In CeCu$_{6-x}$Au$_x$, ${T_{s}}$ decreases linearly with Au concentration and the structural phase transition disappears beyond the critical concentration, $x_{S}$ $\approx$ 0.14, which is close to the magnetic QCP, $x\mathrm{_{QCP}}$ $\approx$ 0.1\cite{grube1999suppression,Löhneysen1990144}. Moreover, one study indicates that the structural transition disappears and magnetic order emerges at nearly the same point in the phase diagram, raising the possibility that a quantum multi-critical point at $x\mathrm{_{QCP}^{S}}$ $\approx$ 0.13 is the origin of the unusual quantum critical behavior in CeCu$_{6-x}$Au$_x$ \cite{robinson2006quantum}. Consequently, further investigation of the influence of the structural phase transition on the unconventional nature of the QCP in CeCu$_{6-x}$Au$_x$ is of interest.

\begin{figure*}
\includegraphics[width=7.0in]{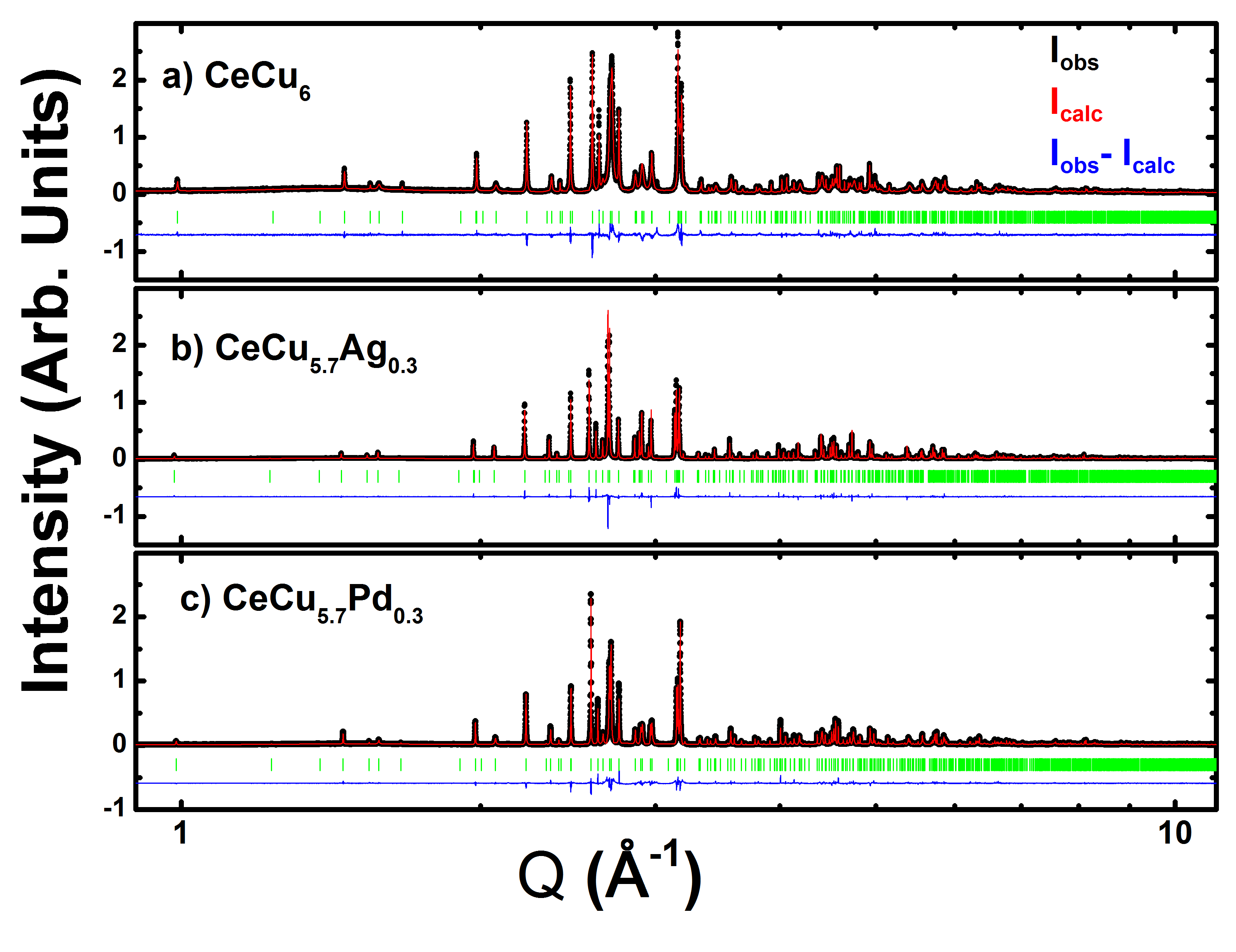}
\caption{(Color online) a) Room temperature x-ray powder diffraction pattern (black dots) plotted along with the Rietveld refinement (red line) for a) CeCu$_{6}$, b) CeCu$_{5.7}$Ag$_{0.3}$ and c) CeCu$_{5.7}$Pd$_{0.3}$. Vertical green marks are the positions of the structural reflections. The diffraction patterns were collected at beamline 11-BM at the Advance Photon Source (APS) of Argonne National Laboratory (ANL). The values of the refined parameters are tabulated in tables \ref{refinement} and \ref{lattice}. The difference ($\mathrm{I_{obs}}$ -- $\mathrm{I_{calc}}$) is offset for clarity. Note: the x-axis (Q) is in a logarithmic scale.}
\label{sync_patt}
\end{figure*}

\begin{figure}[h]
\includegraphics[width=3.3in]{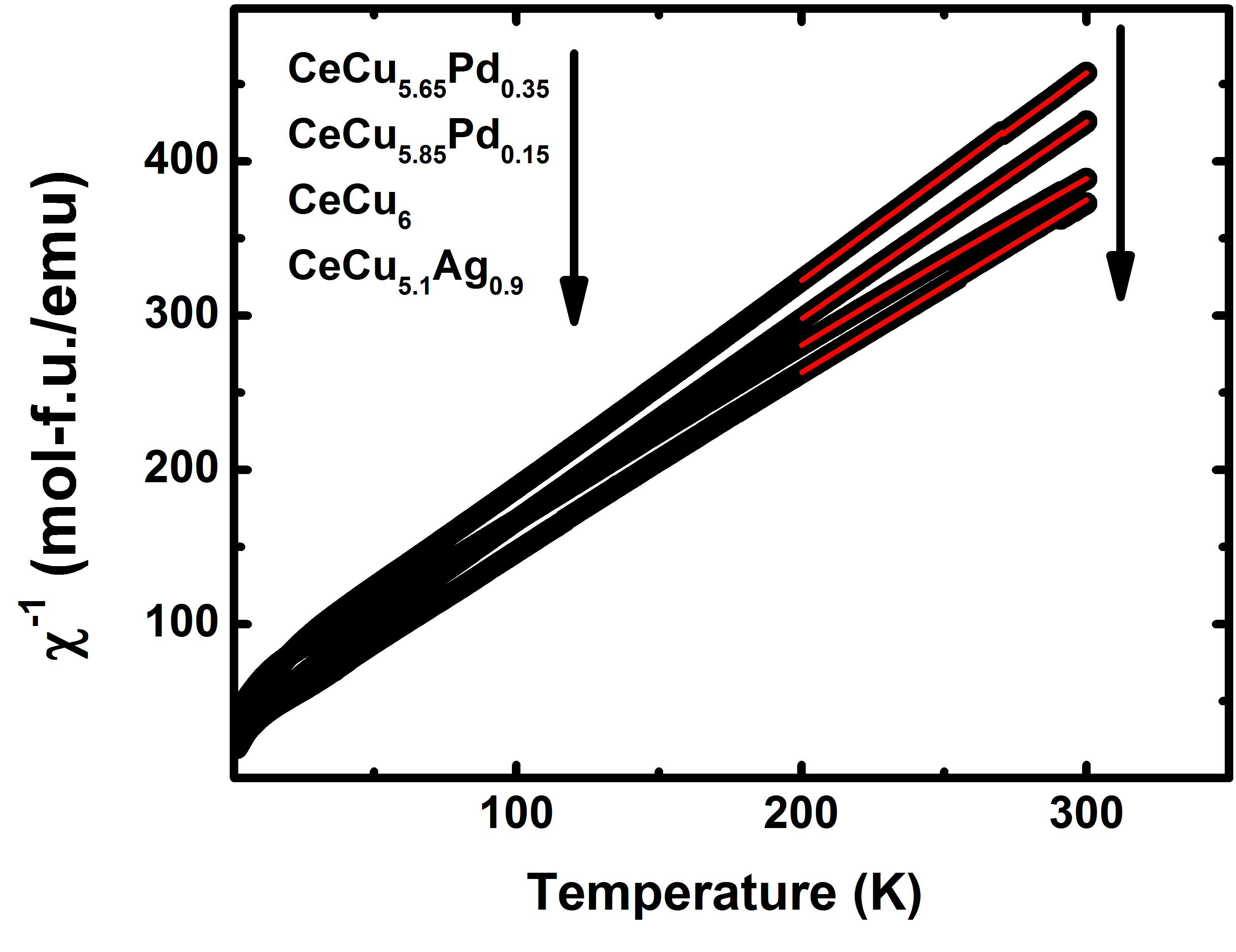}
\caption{(Color online) Inverse susceptibilities of CeCu$_{6-x}$T$_x$ (T = Ag, Pd) showing Curie-Weiss behavior. The red solid line is a Curie-Weiss fit to the data. The effective moments estimated from the Curie-Weiss fits are given in the text. }
\label{susc}
\end{figure}

\begin{table}
\caption{ Lattice parameters and atomic coordinates at room temperature of a) CeCu$_{6}$, b) CeCu$_{5.7}$Ag$_{0.3}$ and c) CeCu$_{5.7}$Pd$_{0.3}$ obtained from Rietveld refinements of the structural model for the synchrotron x-ray measurements (See Fig. \ref{sync_patt}).}

\textbf{a) CeCu$_{6}$}\\
\hspace{5 pt} R$_p$:  18.5 
\hspace{5 pt} R$_{wp}$: 21.0 
\hspace{5 pt} R$_{exp}$: 14.3
\hspace{5 pt} $\chi^2$: 2.1\\
\hspace{5 pt}a = 8.1105(1) ${\mathrm{\AA}}$ \hspace{5 pt}
\hspace{5 pt}b = 5.1000(1) ${\mathrm{\AA}}$ \hspace{5 pt}
\hspace{5 pt}c = 10.1622(2) ${\mathrm{\AA}}$ \hspace{5 pt}
\vspace{0 pt}

\begin{tabular}{ l l l l l l}
\hline 
\hline
Atom   &Wyck.  &   x/a      &      y/b       &      z/c    &  Occupancy\\ 
\hline  
Ce     &4$c$ &0.2595(4)      &0.2500  &0.5646(4)     &1\\
Cu1   &8$d$   &0.0639(5)     \hspace{5pt}  &0.5085(12)\hspace{5pt}   &0.3075(6) \hspace{5pt}    &1\\
Cu2   &4$c$  &0.1469(8)      &0.2500  &0.8597(6)    &1\\
Cu3   &4$c$  &0.3199(8)      &0.2500  &0.2522(5)    &1\\
Cu4   &4$c$  &0.0604(10)      &0.2500  &0.0993(7)    &1\\
Cu5   &4$c$  &0.4063(9)      &0.2500  &0.0134(7)    &1\\
 
\hline
\end{tabular}
\vspace{5 pt}

\textbf{b) CeCu$_{5.7}$Ag$_{0.3}$}\\
\hspace{5 pt} R$_p$:  14.2   
\hspace{5 pt} R$_{wp}$: 16.1   
\hspace{5 pt} R$_{exp}$: 9.43 
\hspace{5 pt} $\chi^2$: 2.93\\

a = 8.1702(1) ${\mathrm{\AA}}$ \hspace{5 pt}
b = 5.0979(1) ${\mathrm{\AA}}$ \hspace{5 pt}
c = 10.2388(1) ${\mathrm{\AA}}$ \hspace{5 pt}
\vspace{0 pt}
\begin{tabular}{l l l l l l} 
\hline
\hline
Atom   &Wyck.  &   x/a      &      y/b        &      z/c     &  Occupancy\\ 
\hline 
Ce    &4$c$ &0.2605(5)     &0.2500     &0.5644(3)     &1 \\
 Cu1 &8$d$   &0.0631(5) \hspace{5pt}    &0.5007(8)  \hspace{5pt}  &0.3119(3) \hspace{5pt}    &1  \\
 Cu2   &4$c$ &0.1459(4)    &0.2500     &0.8593(3)     &0.66(1)\\
 Ag2   &4$c$ &0.1459(4)    &0.2500     &0.8593(3)     &0.34(1)\\
 Cu3  &4$c$  &0.3149(3)    &0.2500     &0.2512(3)     &1\\
 Cu4   &4$c$ &0.0608(4)    &0.2500     &0.1006(4)    &1\\
 Cu5  &4$c$  &0.4029(5)    &0.2500     &0.0146(4)    &1\\

\hline
\end{tabular}

\vspace{5 pt}

\textbf{c) CeCu$_{5.7}$Pd$_{0.3}$}\\

\hspace{5 pt} R$_p$:  15.9 
\hspace{5 pt} R$_{wp}$:  27.9 
\hspace{5 pt} R$_{exp}$: 9.4
\hspace{5 pt} $\chi^2$: 4.82\\
    
a = 8.13412(2) ${\mathrm{\AA}}$ \hspace{5 pt}
b = 5.09530(1) ${\mathrm{\AA}}$ \hspace{5 pt}
c = 10.19498(1) ${\mathrm{\AA}}$ \hspace{5 pt}
\vspace{0 pt}
\begin{tabular}{l l l l l l} 
\hline  
\hline
Atom   &Wyck.  &   x/a      &      y/b        &      z/c     &  Occupancy\\ 
\hline 
Ce    &4$c$   &0.2605(5)        &0.2500(0)        &0.5648(3)          &1\\
 Cu1 &8$d$ &0.0641(6)\hspace{5pt} &0.5062(15)\hspace{5pt} &0.3093(6) \hspace{5pt} &0.96(2)\\
 Pd1   &8$d$   &0.0641(5)        &0.5062(15)        &0.3093(6)          &0.04(2)\\
 Cu2  &4$c$    &0.1443(8)        &0.2500        &0.8583(6)          &0.80(2)\\
 Pd2  &4$c$    &0.1443(8)        &0.2500        &0.8583(6)          &0.20(2)\\
 Cu3  &4$c$    &0.3167(9)        &0.2500        &0.2528(5)          &1\\
 Cu4   &4$c$   &0.0586(10)        &0.2500        &0.0980(8)          &0.94(4)\\
 Pd4  &4$c$    &0.0586(10)        &0.2500        &0.0980(8)          &0.06(4)\\
 Cu5  &4$c$    &0.4030(10)        &0.2500        &0.0158(7)          &1\\
\hline
\end{tabular}

\label{refinement}
\end{table}

Doping CeCu$_6$ with transition metals other than Au offers the opportunity to explore the nature of the QCP in an expanded parameter space where the structural and magnetic degrees of freedom can potentially be decoupled. For example, CeCu$_{6-x}$Ag$_x$\cite{PhysRevB.40.4735,germann1988magnetic,Gango_Ag},  CeCu$_{6-x}$Pd$_x$\cite{Sieck1996325}, CeCu$_{6-x}$Pt$_x$\cite{Sieck1996325} and CeCu$_{6-x}$Sn$_x$\cite{Isnard1999335} all exhibit an antiferromagnetic order that evolves with doping. In the Ag/Pt/Pd-doped systems, deviation from Fermi-liquid behavior is reported at the QCP\cite{Sieck1996325,scheidt1999quantum,PhysRevB.57.R4198}. Furthermore, in  CeCu$_{6-x}$Ag$_x$, thermal expansion measurements indicate that the divergence of the Gr\"uneisen ratio is much weaker than that expected from the HMM model suggesting that the critical behavior is unconventional\cite{kuchler2004}. 

In this paper, we present the first comprehensive neutron and x-ray diffraction investigation of the structural and magnetic properties of the CeCu$_{6-x}$Ag$_x$ and the CeCu$_{6-x}$Pd$_x$ systems. The structural properties were studied using neutron diffraction, resonant ultrasound spectroscopy (RUS), and x-ray diffraction measurements for different compositions of CeCu$_{6-x}$Ag$_x$ and CeCu$_{6-x}$Pd$_x$. Elastic neutron scattering measurements were performed for several members of CeCu$_{6-x}$Ag$_x$ and CeCu$_{6-x}$Pd$_x$ to build a detailed understanding of the evolution of the antiferromagnetic phase with doping. The Neel temperatures ($T_N$) obtained from neutron diffraction measurements in both systems are in agreement with previously published work\cite{Sieck1996325,scheidt1999quantum}. The values of ${T_{s}}$ in CeCu$_{6-x}$Ag$_x$ decrease linearly with Ag-composition, until the structural phase transition disappears at the critical concentration, $x_{S}\approx$ 0.1. In CeCu$_{6-x}$Pd$_x$, no change in the structural transition temperature is observed for 0 $\leq$ $x \leq$ 0.4.

\begin{table*}[!t]
\caption{Structural parameters of CeCu$_{6-x}T_x$ ($T$ = Ag, Pd) extracted from diffraction measurements at room temperature. The lattice parameters were obtained from Rietveld analysis. }
\centering
\begin{tabular}{lllllll}
\hline
\hline
  \hspace{65pt}         & $x$   \hspace{30pt}    & Measurement \hspace{10pt} & a (${\mathrm{\AA}}$)         \hspace{30pt}    & b (${\mathrm{\AA}}$)        \hspace{30pt}      & c (${\mathrm{\AA}}$)         \hspace{30pt}   & Unit Cell Volume (${\mathrm{\AA}}^3$)     \hspace{5pt}  \\ 
\hline
CeCu$_{6}$              &           & x-ray     & 8.1105(1)     & 5.1010(1)     & 10.1622(2)     & 420.43(1)     \\
\hline
                        & 0.035     & neutron   & 8.1215(9)     & 5.0976(6)     & 10.1775(8)     & 421.35(3)     \\
CeCu$_{6-x}$Ag$_{x}$    & 0.1       & neutron   & 8.1266(6)     & 5.0972(3)     & 10.1846(5)     & 421.87(2)     \\
                        & 0.3       & x-ray     & 8.1702(1)     & 5.0979(1)     & 10.2388(1)     & 426.46(1)     \\
\hline
                        & 0.05      & neutron   & 8.1103(6)     & 5.0998(4)     & 10.1649(3)     & 420.43(2)    \\
                        & 0.1       & neutron   & 8.1177(8)     & 5.0996(5)     & 10.1727(9)     & 421.11(3)     \\
CeCu$_{6-x}$Pd$_{x}$    & 0.3       & x-ray     & 8.1341(2)     & 5.0953(1)     & 10.1950(1)     & 422.54(1)     \\
                        & 0.4       & neutron   & 8.1444(8)     & 5.0899(4)     & 10.2042(8)     & 423.01(2)    \\ 
\hline
\end{tabular}

\label{lattice}
\end{table*}

\begin{figure}
\includegraphics[width=3.0in]{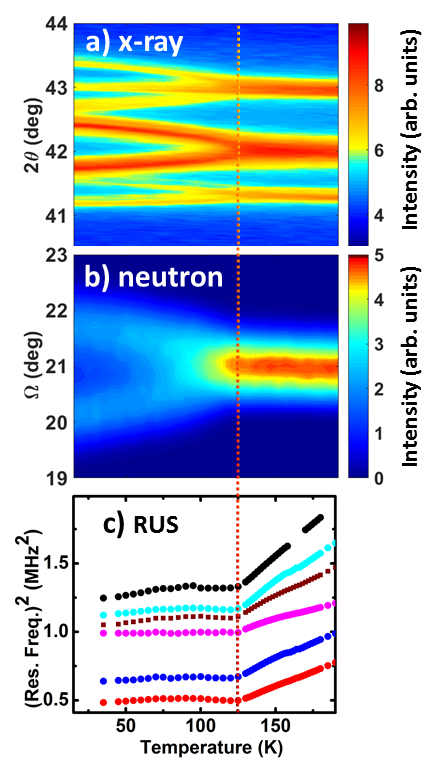}
\caption{(Color online) The structural transition of CeCu$_{5.95}$Ag$_{0.05}$ characterized using a) powder x-ray diffraction with wavelength, $\lambda$ $\approx$ 1.54 ${\mathrm{\AA}}$: The (122), (220) and (221) structural peaks in the orthorhombic phase split in the monoclinic phase as described in the text. Note that peaks indexed as (h 0 l) or (0 k l) in the orthorhombic phase are unaltered, aside from a change of indexing, by the monoclinic distortion.  b) Single crystal neutron diffraction with $\lambda$ $\approx$ 1.003 ${\mathrm{\AA}}$: The structural peak (2 2 0) of orthorhombic phase splits into two structural peaks (2 0 2) and (-- 2 0 2) of monoclinic phase at T$_{S}$. c) RUS: the square of the resonant frequencies change slope at ${T_{s}}$. A vertical line in the plot shows ${T_{s}}$ . All measurements indicate that the structural phase transition takes place in CeCu$_{5.95}$Ag$_{0.05}$ at T$_{S} \approx$ 125(5) K.}
\label{struct_transition}
\end{figure}

\section{\label{sec:level2}Experimental Details}
Single crystals and polycrystalline samples were synthesized for this study. Polycrystalline samples of CeCu$_{6-x}$T$_x$ (T = Ag, Pd) were synthesized by arc melting stoichiometric quantities of Ce (Ames laboratory, purity  =  99.9999\%), Cu (Alpha Aesar, purity  =  99.9999\%), Pd (Alpha Aesar, purity  =  99.9999\%) and Ag (Alpha Aesar, purity  =  99.9999 \%) on a water-cooled copper crucible inside an ultra high purity argon atmosphere. The arc-melted buttons were flipped and remelted no less than four times to ensure homogeneity of the samples. Some samples were annealed at $750\,^{\circ}{\rm C}$ for a week inside a silica tube back filled with argon. No change due to annealing was observed in the room temperature x-ray diffraction pattern or magnetization measurements. Single crystals of CeCu$_{6-x}$T$_x$ (T = Ag, Pd) were grown in a Tri-arc furnace using the Czochralski Technique. The growth was performed on a water-cooled copper hearth under an atmosphere of flowing ultra high purity argon. The crystals were pulled from the melt using a Tungsten seed rod rotating at $\approx$ 30 rev/minat with a speed of $\approx$ 20 mm/hr. 

For analysis of the antiferromagnetic phase, neutron diffraction measurements were performed on several concentrations of CeCu$_{6-x}$Ag$_x$ (single crystals with $x$ = 0.3, 0.35, 0.4, 0.5, 0.75 and polycrystalline samples with $x$ = 0.65, 0.85) and CeCu$_{6-x}$Pd$_x$ (single crystals with $x$ = 0.125, 0.15, 0.4, and polycrystalline samples with $x$ = 0.25, 0.35, 0.4). Approximately 10 g of polycrystalline sample was used for each measurement. The polycrystalline samples were ground inside a glove box and held inside a cylindrical aluminium container loaded in a $^3$He refrigerator. The single crystal measurements were performed with $\approx$ 0.5 g samples. Each single crystal was pre-aligned using the neutron alignment station (CG-1B) at the High Flux Isotope Reactor (HFIR) of Oak Ridge National Laboratory (ORNL). All the samples were measured using the triple-axis spectrometers HB-1A, HB-1, HB-3, and CG-4C at HFIR (ORNL) using fixed incident and final energies of 14.7 meV (HB-1A), 13.5 meV (HB-1), 14.7 meV (HB-3), and 5 meV (CG-4C). A dilution refrigerator provided the sample environment for these measurements.

The structural phase transitions were characterized using RUS, x-ray, and neutron diffraction. High-resolution synchrotron x-ray diffraction patterns were obtained at room-temperature for CeCu$_6$, CeCu$_{5.7}$Ag$_{0.3}$, and CeCu$_{5.7}$Pd$_{0.3}$ from 11-BM at Advance Photon Source(APS) of Argonne National Laboratory (ANL) using x-rays of incident wavelength, $\lambda$ $\approx$ 0.41 ${\mathrm{\AA}}$.  X-ray diffraction patterns on polycrystalline samples of CeCu$_{6-x}$Ag$_x$ ($x$ =  0.015, 0.025, 0.065, 0.075) and CeCu$_{6-x}$Pd$_x$ ($x$ = 0.025, 0.25, 0.3) were collected using a PANalytical X'Pert Pro MPD powder diffractometer. Full patterns were collected at 300 K and 20 K, following which selected peaks were scanned from 10-300 K in 10 K steps for the characterization of the phase transition.  To provide a clear distinction when discussing the details of the crystal structure: a, b, and c are used for the lattice constants in the orthorhombic unit cell whereas a$_m$, b$_m$, and c$_m$ are used for the lattice constants in the monoclinic unit cell.

To further investigate structural properties, neutron diffraction measurements on polycrystalline samples of CeCu$_{6-x}$Ag$_x$ ($x$ = 0, 0.035, 0.1) and CeCu$_{6-x}$Pd$_x$ ($x$ = 0.05, 0.1, 0.4)  were performed using the HB-2A powder diffractometer at HFIR using incident neutrons with wavelength of $\lambda$ = 1.54 ${\mathrm{\AA}}$. In each case, $\sim$5 g of polycrystalline sample was held in a cylindrical vanadium can with  Helium as an exchange gas. The vanadium can was loaded in a top loading closed cycle refrigerator. Diffraction patterns above and below ${T_{s}}$  were collected. A single crystal of CeCu$_{5.95}$Ag$_{0.05}$ was measured using the HB-3A four-circle diffractometer at HFIR with an incident wavelength of $\lambda$ = 1.003 ${\mathrm{\AA}}$. The temperature dependence of the several structural peaks was measured to estimate ${T_{s}}$.

The magnetic susceptibilities of several polycrystalline samples ($x$ = 0.05, 0.3, 0.9 and 1.2 of CeCu$_{6-x}$Ag$_x$, and $x$ = 0.05, 0.15, 0.25, 0.35 and 0.4 of CeCu$_{6-x}$Pd$_x$) were measured using a Quantum Design magnetic property measurement system (MPMS) between 2 K - 300 K with an applied field of 1 kOe. RUS measurements were obtained for  polycrystalline samples of CeCu$_{6-x}$Ag$_x$ ($x$ = 0.025, 0.05, 0.09) using a custom-designed probe in a Quantum Design physical properties measurement system (PPMS). The temperature dependence of the resonances was measured within the frequency range 500-1000 kHz. 

Rietveld analysis of the x-ray and neutron diffraction patterns was performed using the FullProf software package\cite{rodriguez1990fullprof}. A representational analysis was performed using SARAh\cite{sarah} to illuminate symmetry allowed magnetic structures.

\section{Results}
\subsection{\label{sec:levela}Characterization}

Energy-Dispersive x-ray Spectroscopy (EDS) analysis performed on several samples indicate that the samples are homogeneous and the elemental composition is in good agreement with their nominal values. Room-temperature laboratory x-ray diffraction was used for phase identification and as a check of sample purity. Synchrotron x-ray diffraction measurements on the sample compositions noted above were utilized as an additional check of the phase purity of the samples. The only evidence of an impurity phase in the samples studied here is the presence in the synchrotron x-ray diffraction patterns of a single  unidentified peak smaller than 0.25\% of the most intense structural peak of CeCu$_6$. This peak was not observed in the neutron diffraction or the laboratory x-ray measurements.

DC magnetic susceptibility measurements of several samples are shown in Fig. \ref{susc}. A linear dependence of the inverse magnetic susceptibility is observed over a large region of temperature indicating Curie-Weiss behavior. Using the Curie-Weiss relation, $\chi = \chi_0 + \frac{N_A \mu_{eff}^2}{3 k_B (T-\theta_{CW})}$, the effective values of the magnetic moment are estimated from the fits of susceptibility data between 200 K $-$ 300 K. The magnetic moments for different compositions of CeCu$_{6-x}$T$_x$ (T = Ag, Pd) are close to the expected value ($\mu_{eff}$ = 2.54 $\mu_B$) of the Ce$^{3+}$ moment and are 2.42(1) $\mu_B$, 2.43(1) $\mu_B$, 2.67(1) $\mu_B$ and 2.50(1) $\mu_B$  for CeCu$_{6}$, CeCu$_{5.65}$Pd$_{0.35}$, CeCu$_{5.1}$Ag$_{0.9}$ and CeCu$_{5.85}$Pd$_{0.15}$ respectively.

\subsection{\label{sec:levelb}Neutron \& x-ray Diffraction}
The crystal structure of CeCu$_6$ is known to be orthorhombic with space group $Pnma$ at room-temperature\cite{Ruck:se0121,Larson:a03136}. The orthorhombic unit cell consists of one general $8d$ site and five $4c$ sites. The Cerium atoms occupy one of the $4c$ positions  whereas the Cu atoms are distributed among the general $8d$ and four $4c$  sites\cite{Ruck:se0121,Larson:a03136}. This crystal structure is used as a model for the analysis of the diffraction patterns of CeCu$_{6-x}$Ag$_{x}$ and CeCu$_{6-x}$Pd$_{x}$. All the compositions of CeCu$_{6-x}$Ag$_{x}$ and CeCu$_{6-x}$Pd$_{x}$ that we have studied are isomorphous to the parent compound CeCu$_6$ at room temperature. The lattice parameter $b$ slightly decreases with Ag/Pd substitution while the parameters $a$ and $c$ increase along with an overall expansion of the unit cell volume in both systems. The synchrotron x-ray diffraction patterns and the fit obtained from the Rietveld refinements of CeCu$_6$, CeCu$_{5.7}$Ag$_{0.3}$ and CeCu$_{5.7}$Pd$_{0.3}$ are shown in Fig. \ref{sync_patt}, and the results of the fit are summarized in Tables \ref{refinement} and \ref{lattice}.

The coherent neutron scattering lengths of the dopants, Ag (= 5.92 fm) and Pd (= 5.91 fm), are close to that of copper (= 7.72 fm). This low contrast coupled with the small amount of dopants present renders the determination of the site occupancy with neutron scattering inconclusive. Therefore, high-resolution synchrotron x-ray diffraction was used for this purpose. Rietveld refinement of the synchrotron x-ray diffraction measurement indicates that the Ag-atoms in  CeCu$_{6-x}$Ag$_{x}$ are not distributed between different copper sites, but prefer the Cu2 site of the $Pnma$ structure. This is similar to CeCu$_{6-x}$Au$_{x}$ system, where Au-atoms occupy the Cu2 site until the site is fully occupied\cite{Ruck:se0121,mock1994magnetic}. A different situation occurs in  CeCu$_{6-x}$Pd$_{x}$, where the Pd-atoms occupy multiple Cu sites. The analysis of the synchrotron x-ray diffraction pattern shows that the majority of the Pd atoms occupy the Cu2 site and the remaining Pd atoms are distributed on Cu1 and Cu4 sites. The precise value of the Pd-occupancies on all other sites except Cu2 are difficult to determine as the occupancies on these sites are very small and are near the limit of what is possible for this analysis. The Pd-occupancies that give the best fit of the diffraction pattern are shown in Fig. \ref{sync_patt}(c) and are tabulated in the Table \ref{refinement}(b).

\begin{figure}
\includegraphics[width=3.0in]{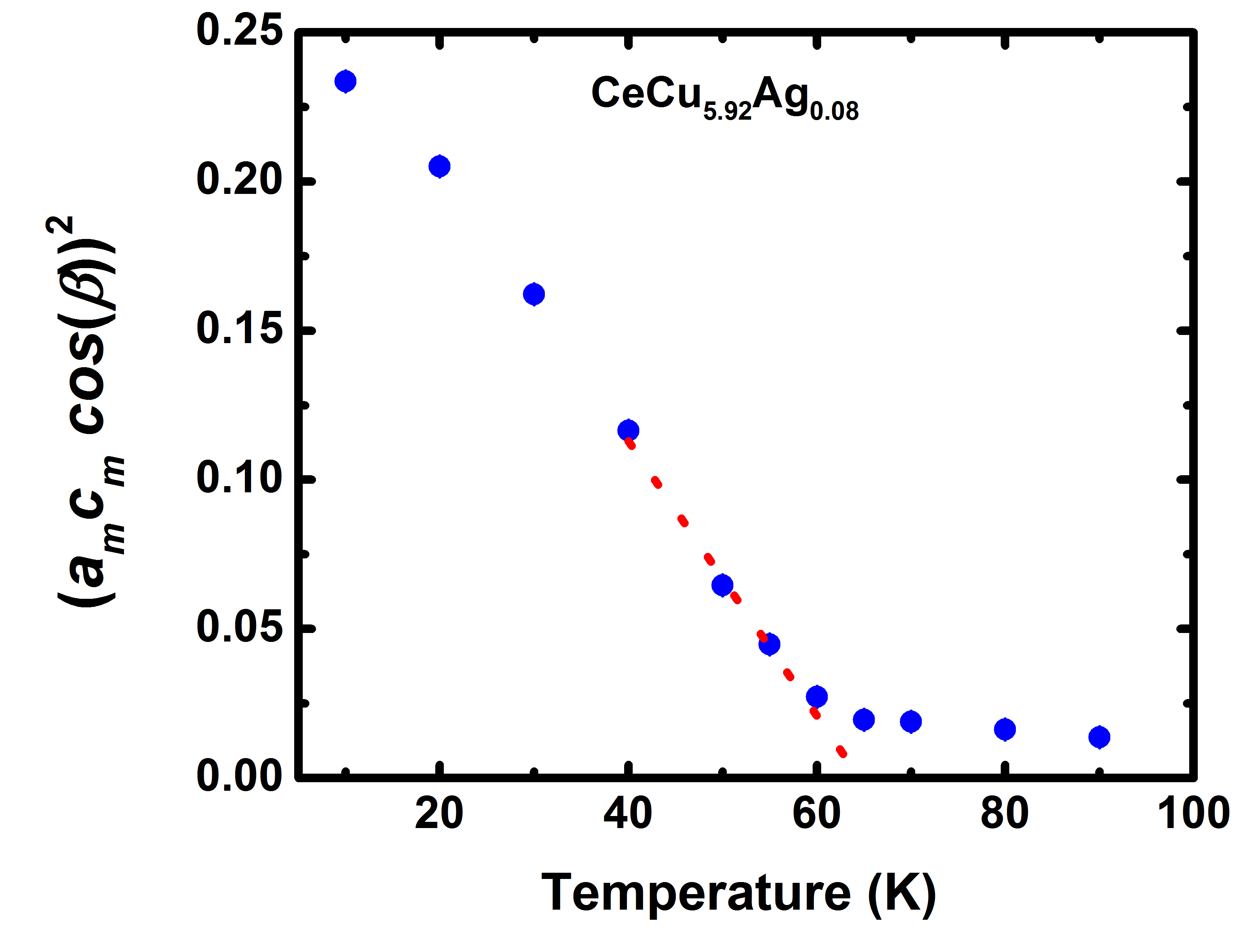}
\caption{ (Color online) Monoclinic order parameter ($(a_mc_m\cos(\beta))^2$) of CeCu$_{5.92}$Ag$_{0.08}$ obtained from neutron diffraction measurements. An extrapolation of the order parameter gives $T_{s}$ = 62(3) K. A continuous change in the monoclinic order parameter is observed near T$_{S}$.}
\label{monoclinic_OP}
\end{figure}

\begin{figure*}[ht]
\includegraphics[width=7.0in]{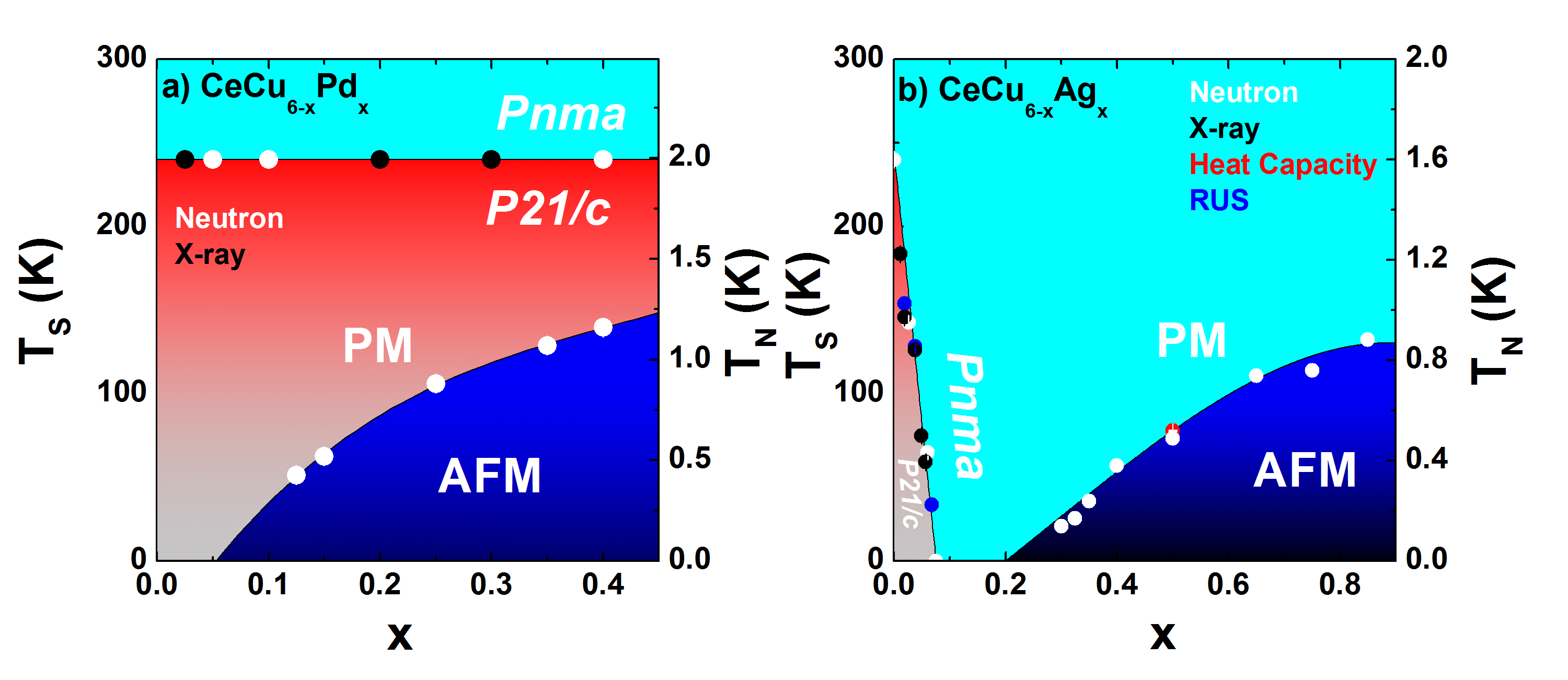}
\caption{(Color online) The phase diagrams of a) CeCu$_{6-x}$Pd$_{x}$ and b) CeCu$_{6-x}$Ag$_{x}$. The concentration for the occurrence of the magnetic QCP is taken to be the same as in \cite{scheidt1999quantum,Sieck1996325} and is $x_\mathrm{QCP} \approx $ 0.2 for CeCu$_{6-x}$Ag$_{x}$ and $x_\mathrm{QCP} \approx $ 0.05 for CeCu$_{6-x}$Pd$_{x}$. In CeCu$_{6-x}$Ag$_{x}$, the termination of the structural phase transition occurs at $x_{S} \approx $ 0.1.   No suppression of the structural phase transition was observed in CeCu$_{6-x}$Pd$_{x}$ for $x$ $\leq$ 0.4. }
\label{phase_diag}
\end{figure*}

\subsection{\label{sec:levelc}Structural Phase Transitions}

Fig. \ref{struct_transition} illustrates the different methods used for the characterization of the structural phase transitions: x-ray diffraction, neutron diffraction, and RUS. To check the consistency in these measurements and the quality of the samples, CeCu$_{5.95}$Ag$_{0.05}$ was grown in both polycrystalline and single crystal forms and measured by all three techniques. A small part of the Czochralski-grown single crystal was measured with neutron diffraction, and different batches of the polycrystalline samples were measured by x-ray diffraction and RUS. The transition temperature (${T_{s}}$) of CeCu$_{5.95}$Ag$_{0.05}$ was estimated to be at ${T_{s}}$ = 125(1) K using neutron diffraction. The RUS and x-ray diffraction measurements of the polycrystalline CeCu$_{5.95}$Ag$_{0.05}$ yield ${T_{s}}$ = 122(1) K and ${T_{s}}$ = 125(5) K respectively. This attests to the consistency of the measurement techniques.

The monoclinic order parameter ($(a_mc_m\cos(\beta))^2$) changes smoothly with temperature near ${T_{s}}$ (Fig. \ref{monoclinic_OP}). The structural peaks that are indexed as (h k l) in space group $Pnma$ split into two structural peaks which are indexed as ( -- k l h) and (k l h) in the monoclinic space group $P2_1/c$. The splitting of the structural peaks at ${T_{s}}$ can be explicitly observed in the temperature dependence of the diffraction pattern as shown in Fig. \ref{struct_transition}(a), where the peaks in $Pnma$ phase split into $P2_1/c$-peaks as (1 2 2) $\rightarrow$ (-- 2 2 1) + (2 2 1), (2 2 0) $\rightarrow$ ( -- 2 0 2) + (2 0 2) and (2 2 1) $\rightarrow$ ( -- 2 1 2) + (2 1 2). The splitting of these peaks is also confirmed by single crystal neutron diffraction measurements, a part of which is shown in Fig. \ref{struct_transition}(b).

RUS measurements are one of the most sensitive ways of characterizing    structural phase transitions. The resonances occur as the natural frequency  of the sample, which is closely related to its elastic properties, matches the incident ultrasonic wave. At ${T_{s}}$, the change in the elastic properties of the sample indicates the occurrence of the structural phase transition. Here, the temperature dependence of the square of the resonant frequency is used to show the structural phase transition. As shown in Fig. \ref{struct_transition}(c), the square of the resonant frequency versus temperature has a constant slope above ${T_{s}}$. The slope of the curve changes continuously at ${T_{s}}$, and the shift in the resonances below ${T_{s}}$ are much weaker as compared to those at above ${T_{s}}$.

The work presented here indicates the structural phase transition from orthorhombic to monoclinic phase in CeCu$_6$ occurs at ${T_{s}}$ $\approx$ 240 K, which is somewhat larger compared to the previous studies\cite{Vrtis1986489,suzukisoft,yamada1987neutron}. The values of ${T_{s}}$ in CeCu$_{6-x}$Ag$_x$ drop linearly with Ag concentration until the structural phase transition disappears above the critical composition, $x_{S}$ $\geq$ 0.1(Fig. \ref{phase_diag}(b)).  For 0.1 $\leq$ $x$ $\leq$ 0.85, no structural phase transition was observed above 4 K. The suppression of the structural phase transition due to doping is analogous to CeCu$_{6-x}$Au$_x$, where ${T_{s}}$ drops in similar fashion and the termination of the structural phase transition occurs at a similar Au-composition, $x_{S}$ $\approx$ 0.14\cite{grube1999suppression,robinson2006quantum}. However, in CeCu$_{6-x}$Pd$_x$, no change in ${T_{s}}$ is observed with Pd-substitution within the range of our investigation, (0 $\leq$ $x$ $\leq$ 0.4). The changes in the transition temperatures with doping in CeCu$_{6-x}$Ag$_x$ and CeCu$_{6-x}$Pd$_x$ are summarized in the phase diagrams presented in Fig. \ref{phase_diag}.

\subsection{First Principles Calculations}

To understand the structural phase transitions, first principles calculations using the planewave code WIEN2K \cite{wien} have been performed.  The generalized gradient approximation (GGA) of Perdew, Burke and Ernzerhof \cite{perdew} was used, with sphere radii for the undoped compound of 2.21 Bohr for Cu and 2.50 for Ce.  For the undoped compound, we used the lattice parameters and angles  of the orthorhombic and monoclinic phases reported by Asano \textit{et al.} \cite{asano}, and relaxed the internal coordinates until forces were less than 2 mRyd/Bohr.  

For the calculations of Ag and Pd doping, several assumptions were made. As detailed in Table \ref{lattice}, in the high temperature orthorhombic phase, for a doping level $x =$ 0.1, the structures for CeCu$_{5.9}$Ag$_{0.1}$ and CeCu$_{5.9}$Pd$_{0.1}$ are very similar, differing in volume by less than 0.2\%. Since even at this doping level the observed low-temperature structures are very different (orthorhombic for Ag doping, monoclinic for Pd doping), we isolated the effects of charge doping from the small structural differences by using the same lattice parameters and internal coordinates for Ag doping and Pd doping.  In each case one of the 24 Cu atoms in the unit cell was replaced by Cu or Pd.  Since there is a substantial site preference for Ag doping in this case we used the Cu2 site for the substitution. For Pd doping, two separate sets of calculations were done, with the Pd atom at the Cu1 and Cu2 sites.  

\begin{table}[]
\caption{The relative energies of several configurations for Ag and Pd doping.  ``SP" refers to a spin-polarized calculation.}
\begin{center}
\begin{tabular}{c|c}
\hline
\hline
Configuration &  $\Delta$ E  (meV/u.c.) \\ \hline
Ag doping --  orthorhombic & 0 \\ 
Ag doping -- orthorhombic SP & -4 \\ 
Ag doping -- monoclinic &  +95 \\ 
Ag doping -- monoclinic SP &  +92 \\ 
Pd doping (Cu2) -  orthorhombic & 0 \\ 
Pd doping (Cu2) - orthorhombic SP & +1 \\ 
Pd doping (Cu2)- monoclinic &  +74 \\ 
Pd doping (Cu2) - monoclinic SP & + 73 \\ 
Pd doping (Cu1) - orthorhombic & 0 \\ 
Pd doping (Cu1) - monoclinic & + 82 \\ \hline
\end{tabular}
\end{center}
\label{dfttab}
\end{table}

\begin{figure}[t]
\includegraphics[width=3.4in]{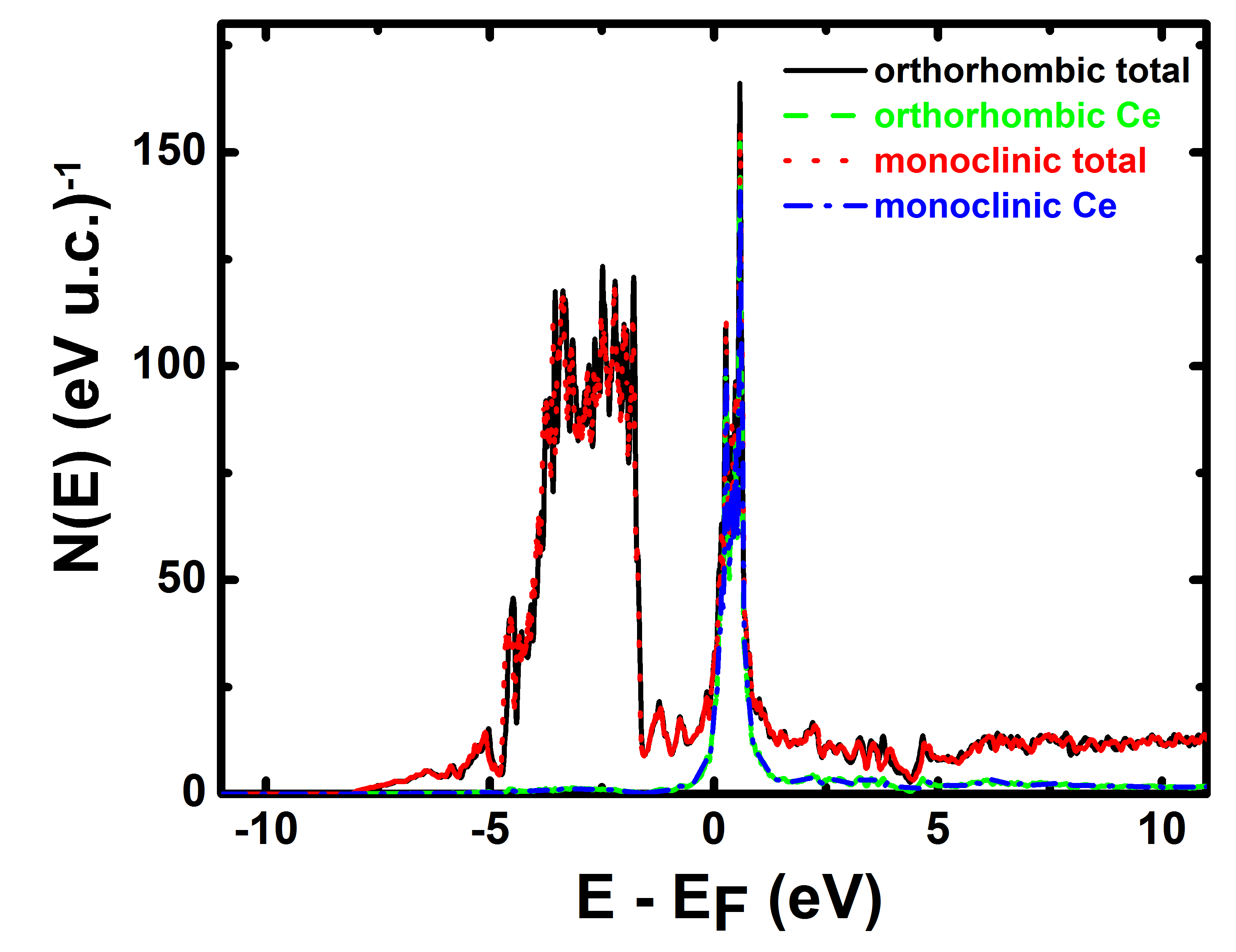}
\caption{(Color online) The calculated density-of-states of CeCu$_{6}$ in the orthorhombic and monoclinic structures.}
\label{dos}
\end{figure}

For CeCu$_6$, the monoclinic structure has an energy 33 meV per unit cell lower than the orthorhombic structure, consistent with experimental observation.  Despite this energy difference, the changes in electronic structure are scant.  Figure \ref{dos} plots the calculated densities-of-states (DOS).  In both cases there is a DOS peak slightly above the Fermi level, attributable to the Ce states, along with a large contribution between 2 and 5 eV beneath E$_F$.  The Fermi level DOS, at 31.53/eV - u.c. for the orthorhombic cell and 30.82/eV - u.c. for the monoclinic, changes by only 2\%.

The relative energies of the doped compounds are presented in table \ref{dfttab}. The non-spin polarized (NSP) orthorhombic state was chosen as the zero of energy.  To make the spin-polarized (SP) calculations tractable, a ferromagnetic Ce configuration was used rather than the actual antiferromagnetism.  The calculations correctly predict the suppression of the monoclinic state with Ag doping - the monoclinic state lies 95 meV higher in energy than the orthorhombic state, and including spin polarization does not appreciably change this result.  The SP orthorhombic state, with Ce moment 0.34 $\mu_B$, is the groundstate, 4 meV beneath the NSP orthorhombic state and well below both monoclinic states.

In contrast to the experiment, the calculations predict a suppression of the monoclinic state in CeCu$_{6-x}$Pd$_x$ system. For Pd doping on the Cu2 site this state is 74 meV above the orthorhombic state, and including spin polarization does not remedy this. A similar result is obtained for doping on the Cu1 site (spin polarization was not checked), with an 82 meV energy difference. 

While the reason for this disagreement is not known, it likely arises from the heavy fermion state, which tends to confound mean-field based density functional theory.  For example, for the base CeCu$_{6}$, our calculated T-linear specific heat coefficient $\gamma$ is 18.2 mJ/mol-K$^{2}$, much smaller than the observed value of between 840 and 1600 mJ/mol-K$^{2}$ \cite{stewart}.  However, since the calculations correctly predict the ground state of CeCu$_6$, and that for Ag doping, one still has hope for the ability of first principles calculations to describe this system.

One plausible way forward would be to explicitly include correlations through an LDA+U approach or an LDA+DMFT approach, as considered by Shim $et$ $al$ \cite{shim} for CeIrIn$_{5}$.  Such approaches would likely tend to yield effective masses in better agreement with experiment, as well as more accurate related quantities such as the relative energies of the orthorhombic and monoclinic states.  LDA+DMFT approaches have been applied, with a fair degree of success, to structural properties of Ce itself \cite{held,zolfl,mcmahan,haule} and we anticipate that similar success can be achieved with LDA+DMFT in studying Ag and Pd doping of CeCu$_{6}$.

\begin{figure*}[ht]
\includegraphics[width=7.0in]{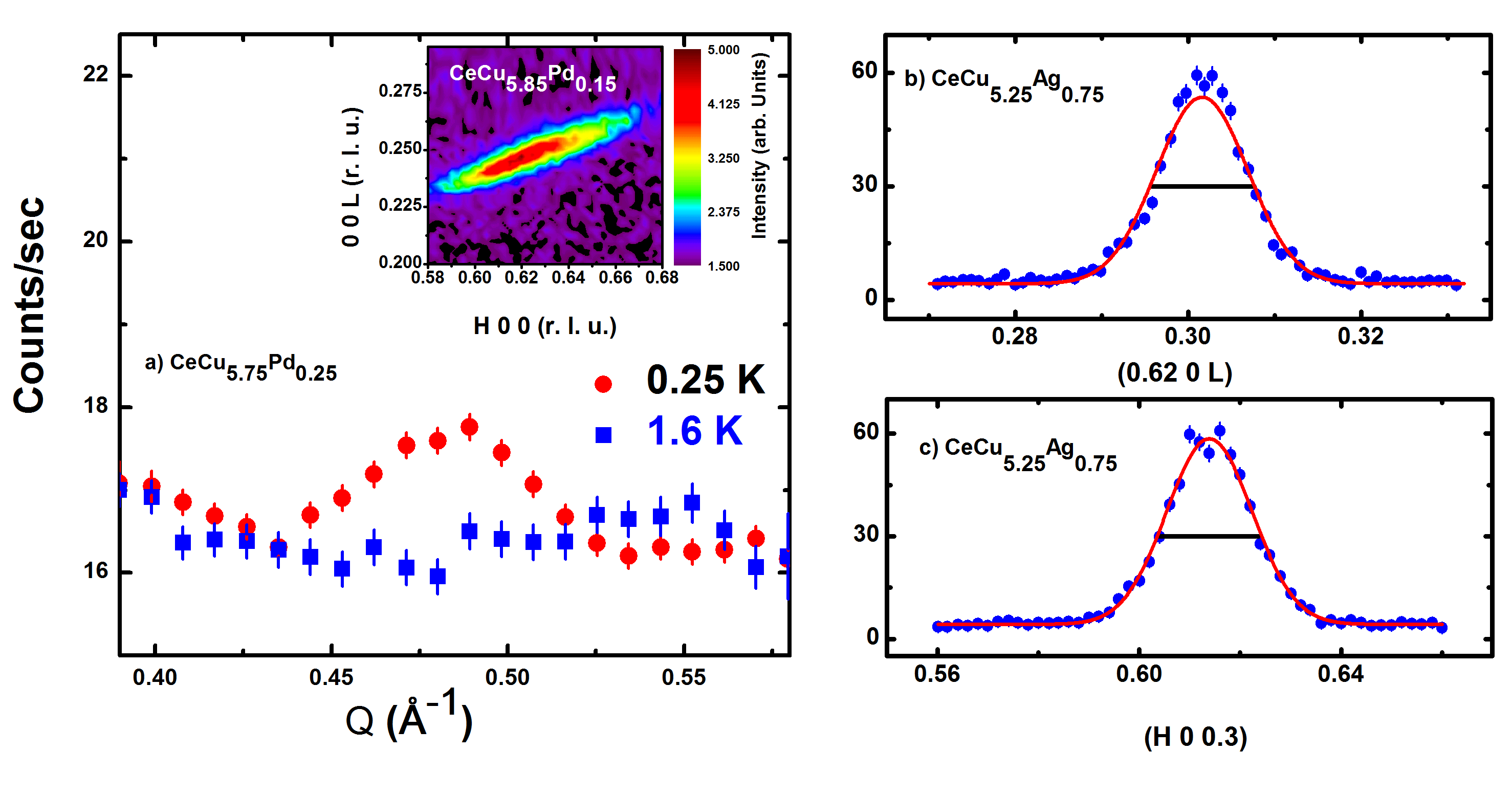}
\caption{(Color online) a) The magnetic reflection at Q = 0.480(3) from a  polycrystalline sample with  a composition of CeCu$_{5.75}$Pd$_{0.25}$. The inset shows the magnetic reflection from a a single crystal of CeCu$_{5.85}$Pd$_{0.15}$ near the wave vector {\bf Q} = (0.6243(1) 0	0.2503(1)). b) and c) Single crystal diffraction measurement of the the (0.62 0 0.3) magnetic reflection of CeCu$_{5.25}$Ag$_{0.75}$ along (0.62 0 L) and (H 0 0.3) at 0.05 K. The red line is a Gaussian fit to the data constrained by the instrumental resolution. The horizontal black lines are the calculated instrumental resolution.}
\label{bragg_peak}
\end{figure*}

\begin{figure*}[ht]
\includegraphics[width=7.0in]{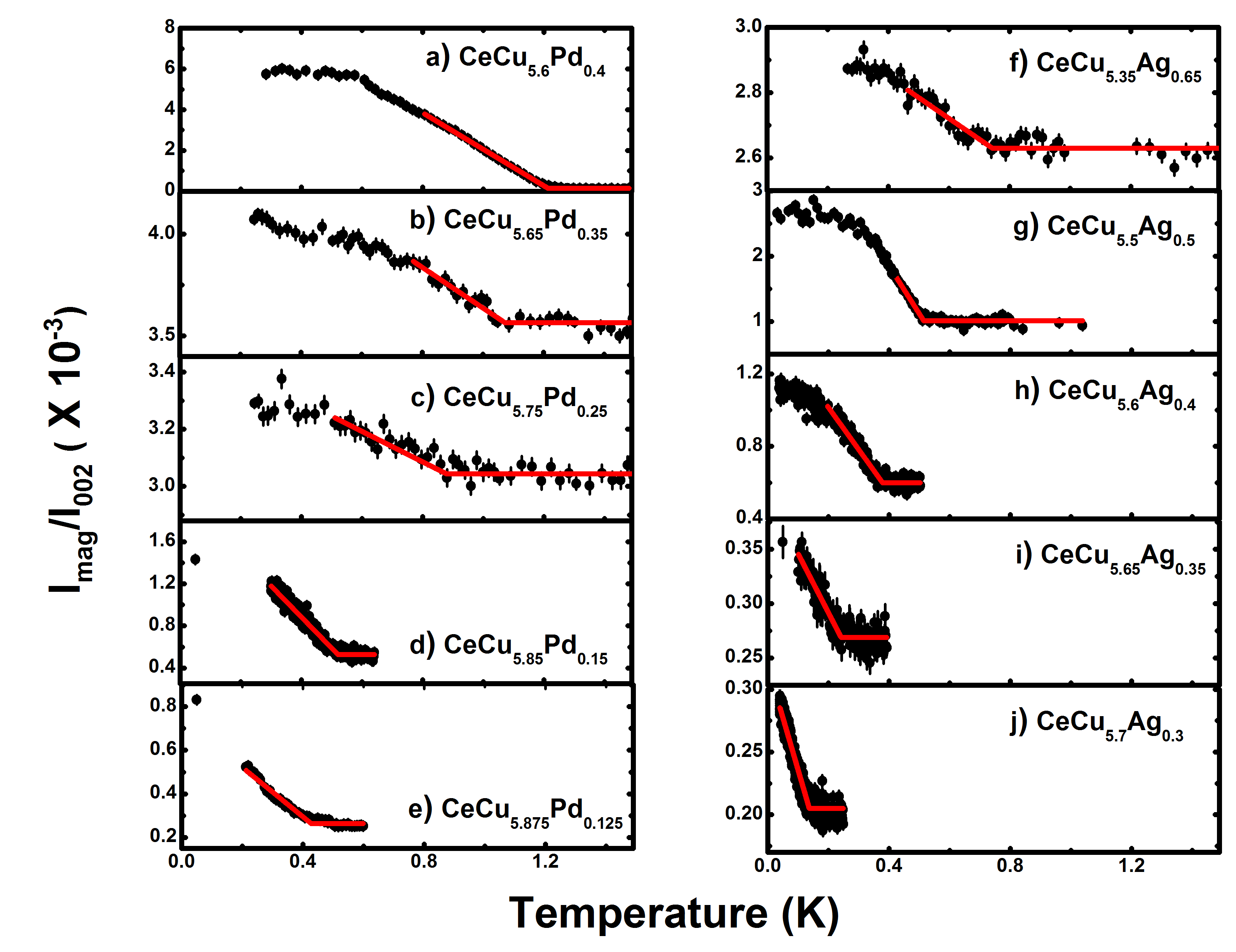}
\caption{(Color online) Temperature dependence of the intensity of the magnetic peak ((0.62 0 0.25) for CeCu$_{6-x}$Pd$_{x}$ and (0.64 0 0.3) for CeCu$_{6-x}$Ag$_{x}$) normalized to the structural (0 0 2) peak in (a-e) CeCu$_{6-x}$Pd$_{x}$ and (f-j) CeCu$_{6-x}$Ag$_{x}$. The red line is a fit to the the data with a power law equation of the form $y = y_0 + A(T-T_N)^{2\beta}$. $\beta$ was fixed to the mean field value of 0.5 and ${T_{N}}$ was allowed to vary in order to estimate the transition temperature. The estimated values of ${T_{N}}$ are a) 1.20(1) K, b) 1.07(2) K, c) 0.88(3) K, d) 0.52(1) K, e) 0.43(1) K, f) 0.74(2) K, g) 0.51(1) K, h) 0.38(1), i) 0.24(1) K, j), and 0.14(1) K. 
 }
\label{magnetic_transition}
\end{figure*}

\begin{figure*}[t]
\includegraphics[width=7in]{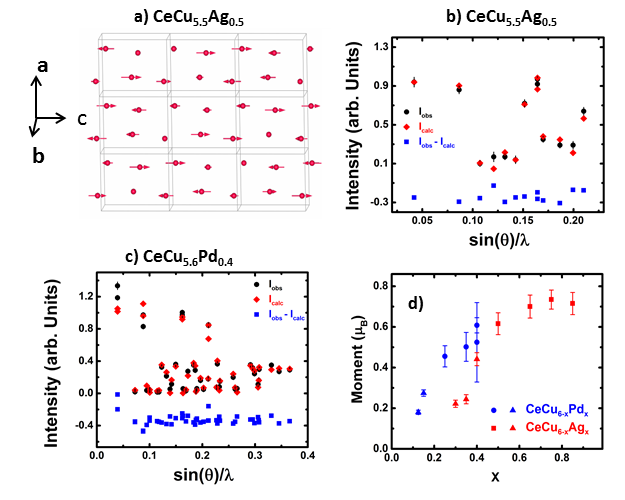}
\caption{(Color online) a) The magnetic structure of CeCu$_{5.5}$Ag$_{0.5}$. Each box represents a structural unit cell that contains four Ce-atoms as indicated by the red spheres. The length and direction of the arrows shows the magnitude and direction of the moment, respectively. All Ce-moments point along along the $c$-axis of the orthorhombic unit cell. The amplitude of the moment is modulated along the wave vector (0.62 0 0.3). The observed and calculated intensities for the magnetic structure is displayed in b) and c) for CeCu$_{5.5}$Ag$_{0.5}$ and  CeCu$_{5.6}$Pd$_{0.4}$ respectively. d) Compositional dependence of the ordered moments in CeCu$_{6-x}$T$_x$ ($T$ = Ag, Pd). The magnetic structure determined for CeCu$_{5.5}$Ag$_{0.5}$ and CeCu$_{5.6}$Pd$_{0.4}$, was assumed to be the same for the remaining members of each series. The QCP occurs at $x = 0.05$ for CeCu$_{6-x}$Pd$_x$ and $x = 0.2$, for CeCu$_{6-x}$Ag$_x$. A point shown as a triangle indicates a composition for which the magnetic moment was not completely saturated and thus the full moment value will be somewhat larger.}
\label{structure}
\end{figure*}

\subsection{\label{sec:leveld}Magnetic Phase Transitions}
Using the information obtained from the study of the polycrystalline samples together with the CeCu$_{6-x}$Au$_x$ literature, magnetic reflections in the (h 0 l) scattering plane were measured in CeCu$_{6-x}$Ag$_x$ and CeCu$_{6-x}$Pd$_x$ (for example see Fig. \ref{bragg_peak}). For simplicity and ease of comparison with the work presented here on CeCu$_{6-x}$Ag$_x$ as well as the extensive work on CeCu$_{6-x}$Au$_x$, we neglect the small monoclinic distortion (less than 2$^\circ$) in CeCu$_{6-x}$Pd$_x$ and use orthorhombic notation to discuss the magnetic properties.

The magnetic Bragg reflections occurs at the points in the reciprocal space that satisfies the condition $
\bf{Q}$ = $\bftau$ $\pm$ $\bf{k_i}$ (i = 1, 2), where $\bftau$ represents a nuclear reciprocal lattice vector and $\bf{k_i}$ are the incommensurate propagation vector, $\bf{k_1}$ = (0.62 0 0.3) , or the symmetry equivalent  $\bf{k_2}$ = (0.62 0 $\mathrm{-}$0.3). The magnetic propagation vector for CeCu$_{6-x}$Ag$_x$ is only weakly dependent on composition. The change in the magnitude of the propagation vector with doping appears to be more pronounced in CeCu$_{6-x}$Pd$_x$ as presented in Table \ref{magnetic_prop}. 

\begin{table}[]
\centering
\caption{Compositional dependence of the magnetic wave vector in CeCu$_{6-x}$Ag$_x$  and CeCu$_{6-x}$Pd$_x$. }
\label{magnetic_prop}
\begin{tabular}{ccc}
\multicolumn{3}{c}{{\bf CeCu$_{6-x}$Ag$_x$}}      \\
\hline
\hline
x     & \hspace{1.5cm} {\bf Q} (h k l) [r.l.u.]   \hspace{1.5cm}             & |Q| (${\mathrm{\AA}^{-1}}$)       \\
\hline
0.85  & ---                    & 0.499(2) \\
0.75  & (0.615(1) 0 0.302(1))   & 0.504(1) \\
0.65  & ---                    & 0.501(2) \\
0.50   & (0.633(1) 0 0.296(1))  & 0.522(1) \\
0.40  & (0.641(1) 0 0.301(1))  & 0.521(1) \\
0.35  & (0.645(2) 0 0.300(1))  & 0.518(1) \\
0.30   & (0.646(1) 0 0.297(1))   & 0.529(1) \\
\multicolumn{3}{c}{{\bf CeCu$_{6-x}$Pd$_x$}}      \\
\hline
\hline
x     & \hspace{1.5cm} {\bf Q} (h k l) [r.l.u.]   \hspace{1.5cm} & |Q| (${\mathrm{\AA}^{-1}}$)     \\
\hline
0.40   & (0.577(1) 0 0.228(1))                    & 0.469(1) \\
0.40   & ---                    & 0.449(3) \\
0.35  & ---                    & 0.466(6) \\
0.25  & ---                    & 0.480(4) \\
0.15  & (0.624(1) 0 0.250(1))  & 0.509(1) \\
0.125 & (0.624(1)  0  0.253(1) & 0.510(1)\\
\hline
\end{tabular}
\end{table}

\begin{table}[t]
\caption{Basis vectors from representational analysis of space group $Pnma$ with ${\bf k}=(0.62,~ 0,~ 0.3)$. The Ce-site is separated into four orbits given by 1: (0.2586, 0.25, 0.5636), 2: (0.2414, 0.75, 0.06360), 3: (0.7414, 0.75, 0.4364), 4: (0.7586, 0.25, 0.9364). The decomposition of the magnetic representation for each of the four orbits is $\Gamma_{Mag}=1\Gamma_{1}^{1}+2\Gamma_{2}^{1}$. Symbols $m_{\|a}$  $m_{\|b}$  $m_{\|c}$ denote the projection of the magnetic moment along the $a$, $b$ and $c$-axis respectively.}
\begin{tabular}{cc|cccccc}
\hline
\hline
\hspace{5pt}IR  &\hspace{10pt}  BV  & \multicolumn{6}{c}{BV components}\\
      &      &             $m_{\|a}$ & $m_{\|b}$ & $m_{\|c}$ &$im_{\|a}$ & $im_{\|b}$ & $im_{\|c}$ \\
\hline
\hspace{5pt}$\Gamma_{1}$ & \hspace{10pt}$\bfpsi_{1}$  &      0 &      1 &      0 &      0 &      0 &      0  \\
                                      
\hline                            
\hspace{5pt}$\Gamma_{2}$ & \hspace{10pt}$\bfpsi_{2}$  &      1 &      0 &      0 &      0 &      0 &      0  \\
             
             & \hspace{10pt}$\bfpsi_{3}$  &      0 &      0 &      1 &      0 &      0 &      0  \\
             
\hline
\end{tabular}

\label{basis_vector_table_1}
\end{table}

The intensities of a structural or a magnetic peak collected with a triple-axis instrument is the convolution of the Bragg intensity with the instrumental resolution function. The software package ResLib\cite{reslib} was utilized to estimate the instrumental resolution and thereby enable the extraction of resolution corrected intensities using the Cooper-Nathans approximation\cite{cooperNathan}. For CeCu$_{5.25}$Ag$_{0.75}$ the calculated instrumental resolution is indicated by the horizontal lines in Fig. \ref{bragg_peak}(b) and \ref{bragg_peak}(c). The instrumental resolution is not isotropic in the scattering plane and results in an elliptically shaped Bragg peak (e.g. inset of Fig. \ref{bragg_peak}(a). The magnetic Bragg peaks observed for all compositions are found to be resolution limited, consistent with the presence of the long range magnetic order.

After the correction for instrumental resolution, magnetic peaks were normalized to the intensity of the nearby structural peak (0 0 2). The resulting order parameter data is used to characterize the magnetic phase transition. The Neel temperatures ($T_N$) are estimated by fitting the power law equation, $y = y_0 + A(T-T_N)^{2\beta}$. At temperatures close to $T_N$, some rounding of the order parameter is observed. The two most likely reasons for this are the presence of critical scattering or a small compositional variation leading to a distribution of T$_N$s. If we attribute the rounding entirely to compositional variation, this indicates a spread of $\Delta T_N \leq$ 20\%, which, with reference to the phase diagram (Fig. \ref{phase_diag}), would imply a compositional variation of $\Delta x \leq$ 0.02. On the other hand, a departure from the behavior of a QCP and the associated expectation of mean field behavior ($\beta$ = 0.5) may be the result of the recovery of classical critical behavior and the presence of critical scattering at temperatures close to T$_N$.  With the currently available data we are unable to distinguish between these two possibilities. Therefore, the value of $\beta$ was restricted to the mean field value of 0.5.  Mean field behavior has previously been observed near the QCP in other heavy fermion systems\cite{Stockert1999376,Jon_CeIn3,pressure_cecu6,Lohn_mag_order}. This is likely due to the change in the effective dimension of the system as the quantum dynamics influences the static critical properties--one of the most prominent distinctions from a counterpart classical phase transition. The fits of the order parameter, for several compositions of CeCu$_{6-x}$Ag$_x$ and CeCu$_{6-x}$Pd$_x$, agree well with the mean-field approximation, as shown in figure \ref{magnetic_transition}.  

To check the consistency with prior work utilizing magnetic susceptibility and heat capacity   to determine $T_N$ \cite{Sieck1996325,PhysRevB.40.4735,kuchler2004}, a heat capacity measurement was performed for CeCu$_{5.5}$Ag$_{0.5}$.  This gives $T_N$ = 0.51(1), which is identical to the value extracted from fitting the order parameter ($T_N$ = 0.50(2)) as well as previously published heat capacity measurements \cite{PhysRevB.40.4735,kuchler2004}. For all other compositions, the estimated values of $T_N$ from the fit of the order parameter are in good agreement with the magnetic susceptibility and heat capacity measurements of previous studies\cite{Sieck1996325,PhysRevB.40.4735,kuchler2004}. The estimated values of ${T_{N}}$ in CeCu$_{6-x}$Ag$_x$ and CeCu$_{6-x}$Pd$_x$ are incorporated in the phase diagrams presented in Fig. \ref{phase_diag}. 

For the wave vector $\bf{k_1}$ = (0.62 0 0.3) and space group $Pnma$, representational analysis indicates the four equivalent Ce positions in the unit cell split into separate orbits as given by 1: (0.2586, 0.25, 0.5636), 2: (0.2414, 0.75, 0.06360), 3: (0.7414, 0.75, 0.4364), 4: (0.7586, 0.25, 0.9364). Two irreducible representations (IR) are found for all four orbits with their basis vectors as listed in Table \ref{basis_vector_table_1}. To understand the magnetic structure in greater detail, we focus on CeCu$_{5.5}$Ag$_{0.5}$ (T$_N$ = 0.51(1) K) and CeCu$_{5.6}$Pd$_{0.4}$ (T$_N$ = 1.20(1) K).  Following the representational analysis, the first IR, $\Gamma_1$, restricts the arrangement of Ce-moments to be parallel to the $b$-axis. Given that the propagation vector indicates a modulation in the $ac$-plane, only a transverse modulation of the magnetic moment is possible under this representation. Structures of this type can discarded as the observed intensities do not match with the calculated intensities. The second IR, $\Gamma_2$, restricts moments to the $ac$-plane but limits the possible magnetic structures to those with a modulation of the magnetic moment amplitude or a cycloidal modulation of the moment direction. However, the cycloidal model does not account well for the intensities of the observed magnetic reflections. The remaining possibility is a sinusoidal modulation of the moment amplitude in the crystallographic $ac$-plane. To simplify the problem, the directions of the Ce-moments were constrained to be same for all four Ce-orbits but the phases were allowed to vary. The best fit under this assumption was obtained when the moments point along the $c$-axis with a magnitude of 0.61(1) $\mu_B$. This yields the magnetic structure shown in the Figure \ref{structure}(a), which is also consistent with the previous studies on CeCu$_{6-x}$Au$_x$ that report a similar structure based on neutron diffraction measurements\cite{lohneysen2006rare,stockert1997247,Okumura1998405,Stockert1999376,hyperscaling}. The comparison between the observed and the calculated intensities is shown in the Figure \ref{structure}(b).

In contrast to CeCu$_{5.5}$Ag$_{0.5}$, the low temperature crystal structure of CeCu$_{5.6}$Pd$_{0.4}$ is monoclinic with space group $P2_1/c$. The representational analysis of CeCu$_{5.6}$Pd$_{0.4}$ using the monoclinic space group along with the corresponding propagation vector ${\bf Q_{monoclinic}}$ = (0 0.23 0.58) indicates that each orbit constitutes a single representation with three real basis vectors along  $a_m$, $b_m$ and $c_m$ axes of the monoclinic structure. The fit of the magnetic structure assuming a sinusoidal modulation of the moment indicates that the moments are pointed along $b_m-$ axis of the monoclinic structure ($c-$axis of the orthorhombic structure \footnote{Note that the structural parameters undergo a cyclic change at the orthorhombic-monoclinic phase transition}), identical to the magnetic structure of CeCu$_{5.5}$Ag$_{0.5}$ described above. The fit of the observed intensity is shown in the figure \ref{structure}c.

The ordered moment obtained from a neutron diffraction measurement is proportional to the square root of the intensity of the magnetic Bragg peak, which is generally scaled to the nuclear reflections to provide an absolute value. Assuming that the magnetic structure doesn't vary with the doping composition, the moments of other compositions were determined from a comparison of the normalized magnetic intensity with respect to CeCu$_{5.5}$Ag$_{0.5}$ and CeCu$_{5.6}$Pd$_{0.4}$. For the compositions near the QCP, a full saturation of the ordered moments was not observed and the real value of the ordered moment is somewhat larger than the estimated values. The variation of the magnetic moment with Ag/Pd-composition is shown in the figure \ref{structure}(c).

\section{\label{sec:level4}Discussion}

It is interesting to view the results presented here on CeCu$_{6-x}$Pd$_x$ and CeCu$_{6-x}$Ag$_x$ in the context of the structural and magnetic properties of the intensively studied CeCu$_{6-x}$Au$_x$ system.  Inelastic neutron scattering measurements at the QCP of CeCu$_{6-x}$Au$_{x}$ show the presence of critical spin fluctuations peaked at {\bf Q} = (0.8 0 0), but present in an extended region of the Brillouin zone in the shape of a butterfly\cite{schroder2000onset,lohneysen2006rare}. Interestingly, the points at the wings of the butterfly correspond to the magnetic ordering wave vectors observed for different Au-compositions of CeCu$_{6-x}$Au$_{x}$ in the magnetically ordered regime\cite{Lohn_mag_order,Löhneysen2000480,Stockert1999376,stockert2010quantum,Löhneysen20022155}. Among all the compositions of CeCu$_{6-x}$Au$_{x}$ that are reported, the composition $x$ = 0.2, studied by neutron diffraction closest to the QCP, is the only one which exhibits short ranged magnetic ordering near the wave vector {\bf Q} = (0.8 0 0) in addition to a long range magnetic ordering that occurs at {\bf Q} = (0.625 0 0.275)\cite{stockert1997247,Lohn_mag_order,Okumura1998405,Stockert1999383}. With a slight increase in Au-composition, for $x$ = 0.3, the short range order disappears and only the long range magnetic ordering is observed at the wave vector {\bf Q} = (0.64 0 0.275)\cite{Lohn_mag_order,Okumura1998405,Stockert1999383}. Upon further alloying with Au, at $x$ = 0.5, the magnetic wave vector exhibits a crossover to {\bf Q} = (0.59 0 0), which stays roughly the same for higher Au-composition \cite{Lohn_mag_order,Okumura1998405,Stockert1999383,Stockert1997250}. However, in CeCu$_{6-x}$Ag$_{x}$ and CeCu$_{6-x}$Pd$_{x}$, our studies find no evidence for short range magnetic order near {\bf Q} = (0.8 0 0). Furthermore, in CeCu$_{6-x}$Ag$_{x}$, the magnetic propagation vector is essentially unchanged for the range of the compositions investigated here suggesting there is no crossover to a propagation vector near $\bf{Q}$ = (0.59 0 0) at large $x$.  

The possibility that the structural degrees of freedom could give rise to a quantum multicritical point in CeCu$_{6-x}$Au$_{x}$\cite{robinson2006quantum} is interesting and worthy of further consideration. In the related system, CeCu$_{6-x}$Ag$_{x}$, the magnetic QCP occurs in the orthorhombic phase, and is well-separated from the termination of the structural phase transition. In CeCu$_{6-x}$Pd$_{x}$, the magnetic QCP occurs in monoclinic phase, and no structural critical point is observed for $x \leq$ 0.4. Despite the distinct behavior of the structural properties of these systems, the magnetic behavior of all three systems is in many ways similar. In particular, the magnetic structures of all these systems are identical to each other: The Ce-moments point along $c-$axis of the orthorhombic unit cell and are modulated with an incommensurate wave-vector. The heat capacity in all three systems appears to exhibit a similar logarithmic divergence at low temperatures \cite{kuchler2004,Löhneysen1996471,Sieck1996325}, indicating the thermal average of the underlining critical fluctuations is independent of the structural properties. These observations suggest that the magnetic QCP is independent of the structural phase transition.

The orthorhombic-monoclinic structural phase transition in CeCu$_6$ is second order in nature\cite{Goto1987309}. We have uncovered no evidence either in the temperature dependence of the monoclinic order parameter $(a_mc_m \cos\beta)^2$ or in the square of the resonance frequencies determined from the RUS measurements that the transition becomes first order with doping (See Fig. \ref{struct_transition}(c) and \ref{monoclinic_OP}).  Thus there appears to be the possibility that a complete suppression of the orthorhombic-monoclinic transition results in a structural QCP. The notion of the structural QCP is still emerging and has attracted recent attention. For example,  the $(\mathrm{Sr},\mathrm{Ca}{)}_{3}{\mathrm{Ir}}_{4}{\mathrm{Sn}}_{13}$ series  appears to exhibit a structural quantum critical point \cite{PhysRevLett.114.097002,PhysRevLett.109.237008}. Since any structural QCP in CeCu$_{6-x}$Au$_{x}$ or CeCu$_{6-x}$Ag$_{x}$ would be complicated by the presence of a magnetic QCP, investigating other CeCu$_6$-derived systems or their non-magnetic analogs, such as LaCu$_{6-x}$Au$_{x}$, may be fertile grounds in which to further probe the concept of a structural QCP.

\section{\label{sec:level5}Conclusions}
In conclusion, we report a comprehensive study of the structural and the magnetic properties of CeCu$_{6-x}$Ag$_{x}$ and CeCu$_{6-x}$Pd$_{x}$. Long range incommensurate magnetic ordering evolves with doping in both systems. The magnetic structure is composed of a sinusoidal modulation of the Ce-moments which are aligned along the $c$-axis of the orthorhombic ($Pnma$) unit cell. The long range magnetic structure as well as the size of the ordered moments determined in CeCu$_{6-x}$Ag$_{x}$ and CeCu$_{6-x}$Pd$_{x}$ are similar to the well known heavy fermion system CeCu$_{6-x}$Au$_{x}$. Yet, these systems exhibit several unique structural and magnetic properties. The magnetic QCP in CeCu$_{6-x}$Ag$_{x}$ occurs in the orthorhombic phase and is well separated from the termination of the structural phase transition. No substantial change in the magnetic wave vector is observed with Ag-composition in CeCu$_{6-x}$Ag$_{x}$. In CeCu$_{6-x}$Pd$_{x}$, the magnetic QCP occurs well within the monoclinic phase. Further investigations of CeCu$_{6-x}$Ag$_{x}$ and CeCu$_{6-x}$Pd$_{x}$ are essential to understand the nature of QCP in these systems.

\begin{acknowledgements}
We acknowledge JM Lawrence for useful discussions, M Suchomel for the technical assistance in synchrotron x-ray diffraction measurements, RE Baumbach and NJ Ghimire for their assistance in the sample preparation. The research at the High Flux Isotope Reactor(ORNL) is supported by the Scientific User Facilities Division, Office of Basic Energy Sciences, U.S. Department of Energy (DOE). Use of the Advanced Photon Source at Argonne National Laboratory was supported by the U. S. Department of Energy, Office of Science, Office of Basic Energy Sciences, under Contract No. DE-AC02-06CH11357. The laboratory XRD work was conducted at the Center for Nanophase Materials Sciences, which is a DOE Office of Science User Facility. HJ, HNL, AFM and DGM acknowledge the support from the U. S. Department of Energy, Office of Science, Basic Energy Sciences, Materials Sciences and Engineering Division.

\end{acknowledgements}

%

\end{document}